\documentclass[twocolumn,english,aps,pra,longbibliography, superscriptaddress,floatfix]{revtex4-1}
\usepackage[T1]{fontenc}
\usepackage[utf8]{inputenc}
\usepackage{color}
\usepackage{xcolor}
\usepackage{amsmath}
\usepackage{graphicx}
\usepackage{natbib}
\usepackage{braket}
\usepackage{amssymb}
\usepackage{dsfont}
\usepackage[normalem]{ulem}
\usepackage[colorlinks=true,urlcolor=blue,citecolor=blue,linkcolor=blue]{hyperref}
\usepackage[ruled,vlined]{algorithm2e}
\usepackage{setspace}
\usepackage{listings}
\usepackage{tikz}
\makeatother

\usepackage{babel}

\begin{document}

\title{Pulse-efficient circuit transpilation for quantum applications on cross-resonance-based hardware}
\author{Nathan Earnest}
\affiliation{IBM Quantum -- IBM T.J. Watson Research Center, Yorktown Heights, New York 10598, USA}
\author{Caroline Tornow}
\affiliation{Institute for Theoretical Physics, ETH Zurich, Switzerland}
\affiliation{IBM Quantum -- IBM Research – Zurich, S\"aumerstrasse 4, 8803 R\"uschlikon, Switzerland}
\author{Daniel J. Egger}
\email{deg@zurich.ibm.com}
\affiliation{IBM Quantum -- IBM Research – Zurich, S\"aumerstrasse 4, 8803 R\"uschlikon, Switzerland}
\date{\today}

\begin{abstract}
We show a pulse-efficient circuit transpilation framework for noisy quantum hardware.
This is achieved by scaling cross-resonance pulses and exposing each pulse as a gate to remove redundant single-qubit operations with the transpiler.
Crucially, no additional calibration is needed to yield better results than a CNOT-based transpilation.
This pulse-efficient circuit transpilation therefore enables a better usage of the finite coherence time without requiring knowledge of pulse-level details from the user. 
As demonstration, we realize a continuous family of cross-resonance-based gates for $SU(4)$ by leveraging Cartan's decomposition.
We measure the benefits of a pulse-efficient circuit transpilation with process tomography and observe up to a 50\% error reduction in the fidelity of $R_{ZZ}(\theta)$ and arbitrary $SU(4)$ gates on IBM Quantum devices.
We apply this framework for quantum applications by running circuits of the Quantum Approximate Optimization Algorithm applied to MAXCUT.
For an 11 qubit non-hardware native graph, our methodology reduces the overall schedule duration by up to 52\% and errors by up to 38\%.
\end{abstract}

\maketitle

\section{Introduction}

Quantum computers have the potential to impact a broad range of disciplines such as quantum chemistry~\cite{Moll2018}, finance \cite{Orus2019, Egger2020}, optimization~\cite{Farhi2014, Egger2020b}, and machine learning~\cite{Biamonte2017, Havlicek2019}.
The performance of noisy quantum computers has been improving as measured by metrics such as the Quantum Volume~\cite{Cross2019, Jurcevic2021} or
the coherence of superconducting transmon-based devices~\cite{Krantz2019, Kjaergaard2020,koch2007charge} which has exceeded $100~\mu{\rm s}$~\cite{Rigetti2012, IBMQuantum}.
To overcome limitations set by the noise, several error mitigation techniques such as readout error mitigation~\cite{Bravyi2020, Barron2020a} and Richardson extrapolation~\cite{Temme2017, Kandala2018} have been developed.
Gate families with continuous parameters further improve results~\cite{LaCroix2020, Foxen2020,gokhale2020optimized} as they require less coherence time than circuits in which the CNOT is the only two-qubit gate.
Aggregating instructions and optimizing the corresponding pulses, using e.g. gradient ascent algorithms such as GRAPE~\cite{Khaneja2005}, reduces the duration of the pulse schedules~\cite{Shi2019}.
However, such pulses require calibration to overcome model errors~\cite{Egger2014, Wittler2020} which typically needs closed-loop optimization~\cite{Kelly2014, Werninghaus2021} and sophisticated readout methods~\cite{Rol2017, Werninghaus2020}. 
This may therefore be difficult to scale as calibration is time consuming and increasingly harder as the control pulses become more complex.
Some of these limitations may be overcome with novel control methods~\cite{Machnes2018}.

Since calibrating a two-qubit gate is time-consuming, IBM Quantum~\cite{IBMQuantum} backends only expose a calibrated CNOT gate built from echoed cross-resonance pulses~\cite{Chow2011, Sheldon2016} with rotary tones~\cite{Sundaresan2020}.
Quantum circuit users must therefore transpile their circuits to CNOT gates which often makes a poor usage of the limited coherence time.
With the help of Qiskit pulse~\cite{Mckay2018, Alexander2020} users may extend the set of two-qubit gates \cite{Garion2020, Oomura2021, Heya2021}.
Such gates can in turn generate other multi-qubit gates more effectively than when the CNOT gate is the only two-qubit gate available~\cite{Oomura2021}.
However, creating these gates comes at the expense of additional calibration which is often impractical on a queue-based quantum computer.
Furthermore, only a limited number of users can access these benefits due to the need for an intimate familiarity with quantum control.
In Ref.~\cite{Stenger2020} the authors show a pulse-scaling methodology to create the control pulses for the continuous gate set $R_{ZX}(\theta)$ which they leverage to create $R_{YX}(\theta)$ gates and manually assemble into pulse schedules.
Crucially, the scaled pulses improved gate fidelity without the need for any extra calibration.

Here, we extend the methodology of Ref.~\cite{Stenger2020} to arbitrary $SU(4)$ gates and show how to make pulse-efficient circuit transpilation available to general users without having to manipulate pulse schedules.
In Sec.~\ref{sec:rzx} we review the pulse-scaling methodology of Ref.~\cite{Stenger2020} and carefully benchmark the performance of $R_{ZZ}$ gates.
Next, in Sec.~\ref{sec:su4}, we leverage this pulse-efficient gate generation to create arbitrary $SU(4)$ gates which we benchmark with quantum process tomography~\cite{Mohseni2008, Bialczak2010}.
In Sec.~\ref{sec:transpilation} we show how pulse-efficient gates can be included in automated circuit transpiler passes.
Finally, in Sec.~\ref{sec:demo} we demonstrate the advantage of our pulse-efficient transpilation by applying it to the Quantum Approximate Optimization Algorithm (QAOA)~\cite{Farhi2014}.

\section{Scaling hardware-native cross-resonance gates\label{sec:rzx}}

We consider an all-microwave fixed-frequency transmon architecture that implements the echoed cross-resonance gate~\cite{Sheldon2016}. 
A two-qubit system in which a control qubit is driven at the frequency of a target qubit evolves under the time-dependent cross-resonance Hamiltonian $H_\text{cr}(t)$. 
The time-independent approximation of $H_\text{cr}(t)$ is
\begin{align}
    \Bar{H}_{cr}=\frac{1}{2}\left(Z\otimes B + I\otimes C\right)
\end{align}
where $B=\omega_{ZI}I+\omega_{ZX}X+\omega_{ZY}Y+\omega_{ZZ}Z$ and $C=\omega_{IX}X+\omega_{IY}Y+\omega_{IZ}Z$.
Here, $X$, $Y$, and $Z$ are Pauli matrices, $I$ is the identity, and $\omega_{ij}$ are drive strengths.
An echo sequence \cite{Sheldon2016} and rotary tones \cite{Sundaresan2020} isolate the $ZX$ interaction which ideally results in the unitary $R_{ZX}(\theta)=\exp\{-i\theta ZX/2\}$.
The rotation angle $\theta$ is $t_{cr}\omega_{ZX}(\bar{A})$ where $t_{cr}$ is the duration of the cross-resonance drive.
The drive strength $\omega_{ZX}$ has a non-linear dependency on the average drive-amplitude $\bar{A}$ as shown by a third-order approximation of the cross-resonance Hamiltonian~\cite{Magesan2020, Alexander2020}. 

IBM Quantum systems expose to their users a calibrated CNOT gate built from $R_{ZX}(\pi/2)$ rotations implemented by the echoed cross-resonance gate.
The pulse sequence of $R_{ZX}(\pi/2)$ on the control qubit is ${\rm CR}(\pi/4)X{\rm CR}(-\pi/4)X$.
Here, ${\rm CR}(\pm \pi/4)$ are flat-top pulses of amplitude $A^*$, width $w^*$, and Gaussian flanks with standard deviations $\sigma$, truncated after $n_\sigma$ times $\sigma$.
Their area is $\alpha^*=|A^*|[w^*+\sqrt{2\pi}\sigma \mathrm{erf}(n_\sigma)]$ where the star superscript refers to the parameter values of the calibrated pulses in the CNOT gate.
During each ${\rm CR}$ pulse rotary tones are applied to the target qubit to help reduce the magnitude of the undesired $\omega_{IY}$ interaction.
We can create $R_{ZX}(\theta)$-rotations by scaling the area of the $\rm CR$ and rotary pulses following $\alpha(\theta)=2\theta\alpha^*/\pi$ as done in Ref.~\cite{Stenger2020}.
To create a target area $\alpha(\theta)$ we first scale $w$ to minimize the effect of the non-linearity between the drive strength $\omega_{ZX}(\bar A)$ and the pulse amplitude.
When $\alpha(\theta)<|A^*|\sigma\sqrt{2\pi}\mathrm{erf}(n_\sigma)$ we set $w=0$ and scale the pulse amplitude such that $|A(\theta)|=\alpha(\theta)/[\sigma\sqrt{2\pi}\mathrm{erf}(n_\sigma)]$.

\begin{figure}
    \centering
    \includegraphics[width=\columnwidth]{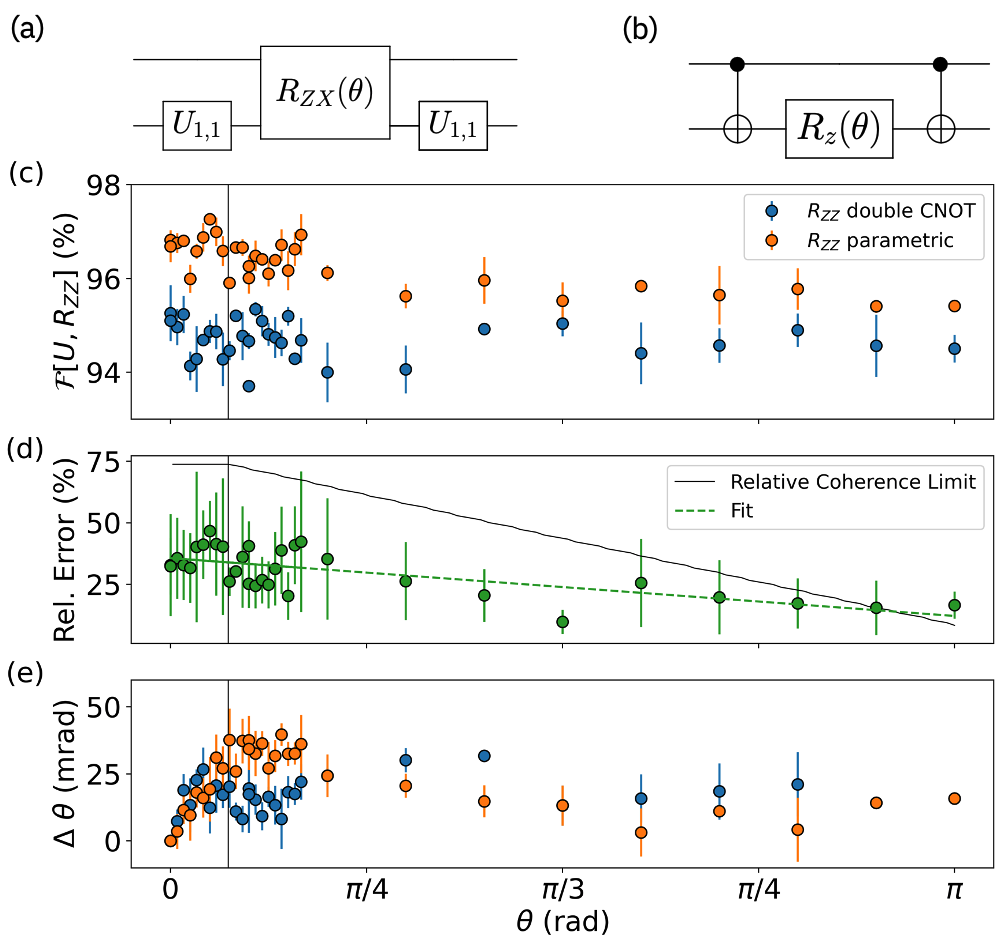}
    \caption{$R_{ZZ}(\theta)$ characterization for qubits one and two on \emph{ibmq\_mumbai}. 
    (a) Double-CNOT benchmark.
    (b) Continuous gate implementation where $U_{1,1}=R_Z(\pi/2)\sqrt{X}R_Z(\pi/2)$.
    Here, $R_{ZX}(\theta)$ is a scaled cross-resonance pulse with a built-in echo.
    (c) Gate fidelity $\mathcal{F}[U_\text{meas}, R_{ZZ}(\theta)]$ of the double-CNOT implementation (blue) and the scaled cross-resonance pulses (orange).
    The vertical line indicates the angle at which $w=0$. 
    (d) The relative error between the two implementations (green dots), and the theoretical expectations for a coherence limited gate (solid black line).
    (e) The deviation angle $\Delta \theta=\theta-\theta_\text{max}$ corresponding to the data in (c) that achieves the maximum gate fidelity $\mathcal{F}[U_\text{meas}, R_{ZZ}(\theta_{\text{max}} )]$.}
    \label{fig:QPT_angle_dev}
\end{figure}

We investigate the effect of the pulse scaling methodology with quantum process tomography by carefully benchmarking scaled $R_{ZZ}(\theta)$ gates, see Fig.~\ref{fig:QPT_angle_dev}(a), with respect to the double-CNOT decomposition, see Fig.~\ref{fig:QPT_angle_dev}(b).
We measure the process fidelity $\mathcal{F}[U_\text{meas}, R_{ZZ}(\theta)]$ between the target gate $R_{ZZ}(\theta)$ and the measured gate $U_\text{meas}$.
To determine $U_\text{meas}$ we prepare each qubit in $\ket{0}$, $\ket{1}$, $(\ket{0}+\ket{1})/\sqrt{2}$, and $(\ket{0}+i\ket{1})/\sqrt{2}$ and measure in the $X$, $Y$, and $Z$ bases.
Two qubit process tomography therefore requires a total of 148 circuits for each angle of interest which includes four circuits needed to mitigate readout  errors~\cite{Bravyi2020, Barron2020a}.
The scaled pulses consistently have a better fidelity than the double CNOT benchmark as demonstrated by the data gathered on \emph{ibmq\_mumbai} with qubits one and two, see Fig.~\ref{fig:QPT_angle_dev}(c).
Appendix~\ref{sec:appendix_additional_data} shows key device parameters and additional data taken on other IBM Quantum devices which illustrates the reliability of the methodology.
The relative error reduction of the measured gate fidelity correlates well to the relative error reduction of the coherence limited average gate fidelity~\cite{Horodecki1999, Magesan2011, Sundaresan2020}, see Fig.~\ref{fig:QPT_angle_dev}(d) and details in Appendix~\ref{sec:appendix_fidelity_limit}.
We therefore attribute the error reduction to the shorter schedules as they use less coherence time.

In addition to the gate fidelity, we compare the deviation $\Delta\theta$ from the target angle of both implementations of the $R_{ZZ}(\theta)$ rotation. 
The deviation $\Delta\theta$ is the difference between the target rotation angle $\theta$ and the angle $\theta_{\text{max}}$ which satisfies $\mathcal{F}[U_\text{meas}, R_{ZZ}(\theta_\text{max})]\geq \mathcal{F}[U_\text{meas}, R_{ZZ}(\theta')]~\forall~\theta'$.
Since the $R_Z(\theta)$ is virtual~\cite{Mckay2017} the implementation with two CNOT gates does not depend on the desired target angle, see Fig.~\ref{fig:QPT_angle_dev}(e). 
However, the scaled gate has two competing non-linearities: an expected non-linearity from the amplitude scaling and an unexpected one from scaling the width.
As the width is scaled down, the angle deviation increases from $\sim\!\!10~{\rm mrad}$ to $\sim\!\!35~{\rm mrad}$.
Once the amplitude scaling begins, a non-linearity arises which reduces the deviation angle of the scaled gates. 
At $\alpha(\theta) \approx |A^*|\sigma\sqrt{2\pi}\mathrm{erf}(n_\sigma)/2$ the angle deviation of the scaled gates once again matches the deviation of the benchmark within the measured standard deviation.

\section{Creating arbitrary SU(4) gates\label{sec:su4}}

\begin{figure}[htbp]
  \includegraphics[width=\columnwidth]{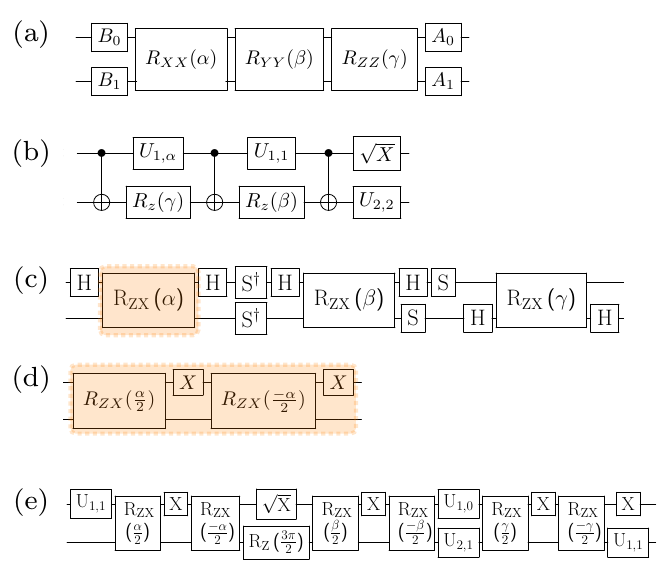}
  \caption{Cartan's $KAK$ decomposition. (a) Circuit representation of the $KAK$ decomposition of a two-qubit gate $U\in SU(4)$ with $k_1 = (A_1 \otimes A_0)$ and $k_2 = (B_1 \otimes B_0)$. (b) Circuit in (a) without $k_{1,2}$ and decomposed into three CNOT gates and transpiled to the basis gates $(R_Z(\theta), \sqrt{X}, \mathrm{CNOT})$. (c) Circuit in (a) decomposed into the hardware-native $R_{ZX}$ gates. Here, each $R_{ZX}$ gate has a built-in echo as shown in (d). 
  Transpiling circuit (c) to the basis $(R_Z(\theta), \sqrt{X}, R_{ZX}(\theta))$ with the echoes exposed to the transpiler results in the pulse-efficient circuit shown in (e) where the scaled $R_{ZX}$ gates do not have an echo.
  We replaced $R_Z(n\pi/2)\sqrt{X}R_Z(m\pi/2)$ with $U_{n,m}$ and $U_{1,\alpha} = R_Z(\pi/2)\sqrt{X}R_Z(\alpha)$ to shorten the notation.
  \label{fig:KAK}}
\end{figure}

\begin{figure}[tbp!]
    \centering
    \includegraphics[width=\columnwidth, clip, trim=20 0 0 0 ]{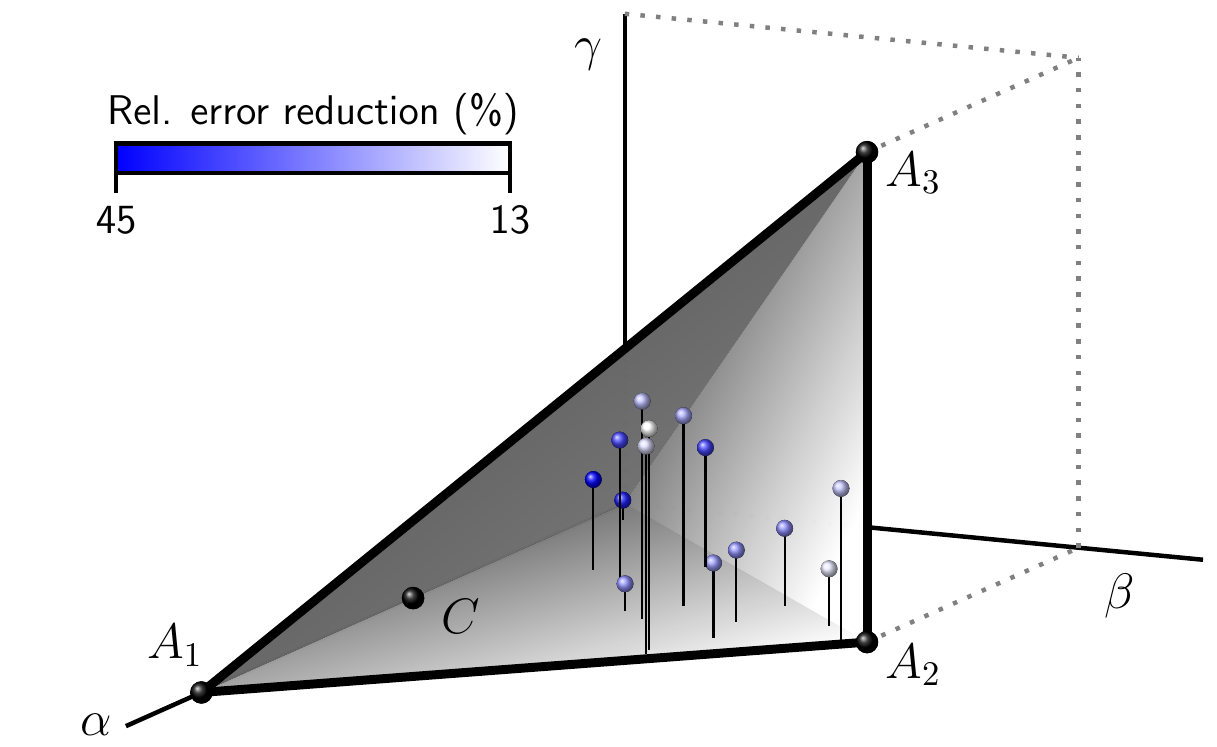}
    \caption{Weyl Chamber of $SU(4)$. The coordinates of the chamber are $O=(0,0,0)$, $A_1=(\pi,0,0)$, $A_2=(\frac{\pi}{2}, \frac{\pi}{2}, 0)$, and $A_3=(\frac{\pi}{2}, \frac{\pi}{2}, \frac{\pi}{2})$.
    $C$ corresponds to the $\rm CNOT$ gate.
    The blue dots represent the data from Fig.~\ref{fig:SU4} taken on \emph{ibmq\_mumbai}.
    }
    \label{fig:weyl_chamber}
\end{figure}

We now generalize the results from Sec.~\ref{sec:rzx}.
Cartan's decomposition of an arbitrary two-qubit gate $U\in SU(4)$ is $U = k_1 A k_2$ which we refer to as Cartan's $KAK$ decomposition~\cite{Khaneja2001}.
Here $k_1$ and $k_2$ are local operations, i.e. $k_{1,2} \in SU(2) \otimes SU(2)$, and $A = e^{i\boldsymbol{k}^T\cdot\boldsymbol{\Sigma}/2} \in SU(4) \setminus SU(2) \otimes SU(2)$ is a non-local operation with $\boldsymbol{\Sigma}^T=(XX, YY, ZZ)$~\cite{Zhang2003,Tucci2005, Byron2008}, see Fig.~\ref{fig:KAK}(a).
The non-local term is defined by the three angles $\boldsymbol{k}^T=(\alpha, \beta, \gamma)\in\mathbb{R}^{3}$ satisfying $\alpha+\beta+\gamma\leq 3\pi/2$ and $\pi\geq\alpha\geq\beta\geq\gamma\geq0$.
Geometrically, the $KAK$ decomposition is represented in a tetrahedron known as the Weyl chamber in the three-dimensional space, see Fig.~\ref{fig:weyl_chamber}.
Every point $(\alpha, \beta, \gamma)$ in the Weyl chamber (except in the base) defines a continuous set of two-qubit gates equivalent up to single-qubit rotations~\cite{Zhang2003}.
For instance, the point $(\frac{\pi}{2}, 0, 0)$, labeled as $C$ in Fig.~\ref{fig:weyl_chamber}, corresponds to the local equivalence class of the CNOT gate, and the point $(\frac{\pi}{2},\frac{\pi}{2},\frac{\pi}{2})$, labeled as $A_3$, represents the SWAP gate.

\begin{figure}[htbp]
\centering
  \includegraphics[width=1.\columnwidth]{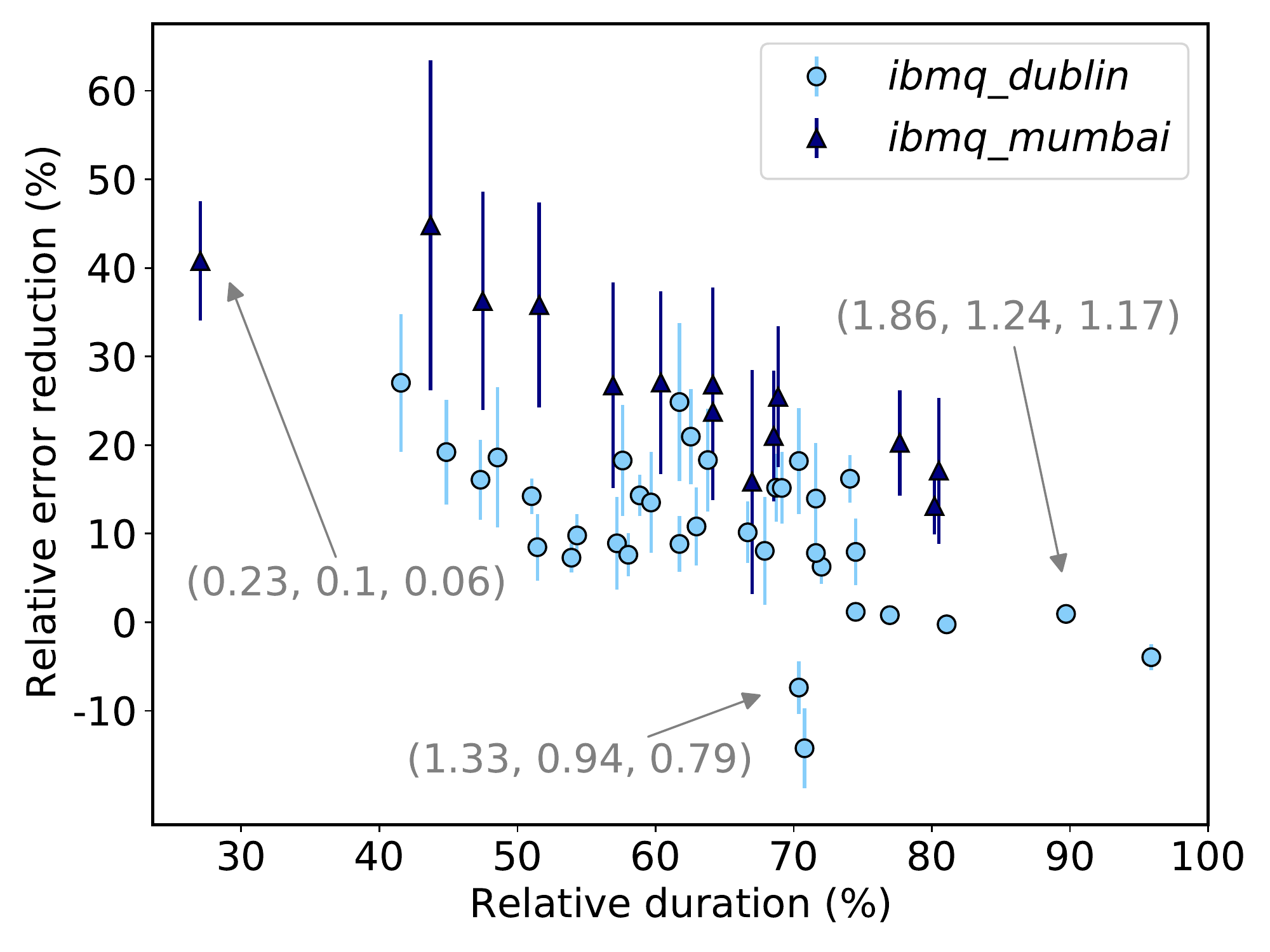}
  \caption{Gate error reduction of the pulse-efficient $SU(4)$ gates relative to the three CNOT benchmark for random angles in the Weyl chamber measured on  \emph{ibmq\_dublin}, qubits one and two (light blue circles), and  \emph{ibmq\_mumbai}, qubits 19 and 16 (dark blue triangles).
  The $x$-axis is the duration of the pulse-efficient $SU(4)$ gates relative to the three CNOT benchmark.
  The angles of three gates are indicated in parenthesis as example.
  \label{fig:SU4}}
\end{figure}

\begin{figure*}[htbp!]
  \includegraphics[width=\textwidth]{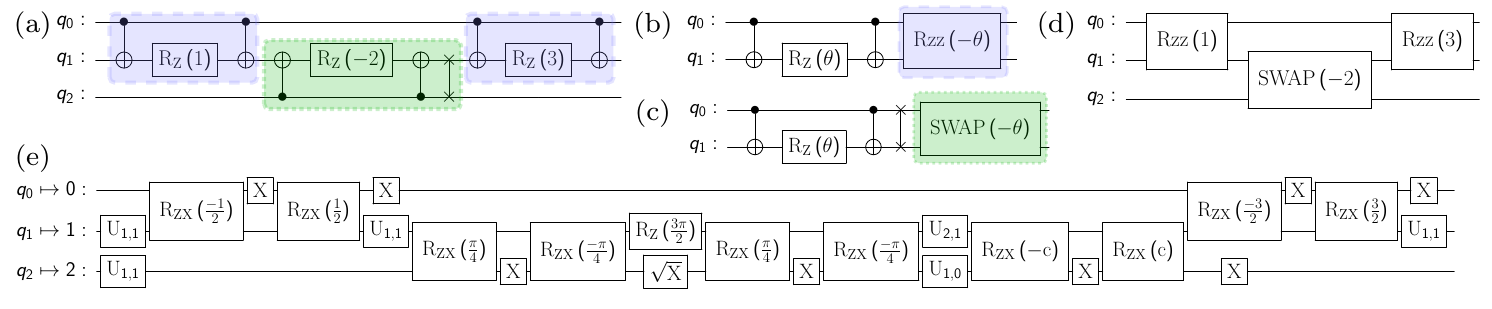}
  \caption{Pulse-efficient transpilation example. (a) Circuit of the cost operator for a QAOA circuit implemented on three qubits connected in a line. (b) and (c) Templates of the $R_{ZZ}$ and phase-swap gates, respectively.
  Here, $R_{ZZ}(\theta)$ and $\mathrm{SWAP}(\theta)$ hold the rules with which to decompose them into the hardware-native $R_{ZX}$ gates.
  (d) Circuit resulting from the template matching of (b) and (c) performed on circuit (a).
  (e) Circuit resulting from a transpilation of (d) which uses the decompositions rules of $R_{ZZ}(\theta)$ and $\mathrm{SWAP}(\theta)$ into $R_{ZX}$.
  To shorten the circuit figure we replaced $R_Z(n\pi/2)\sqrt{X}R_Z(m\pi/2)$ with $U_{n,m}$ and $c=0.215$.
  \label{fig:tranpiler_example}}
\end{figure*}

Since the rotations generated by $XX$, $YY$, and $ZZ$ are locally equivalent to rotations generated by $ZX$ we decompose the non-local $e^{i\boldsymbol{k}^T\cdot\boldsymbol{\Sigma}/2}$ term into a circuit with three $R_{ZX}$ rotations, see Fig.~\ref{fig:KAK}(c).
We shorten the total duration of the circuit by exposing the echo in the cross-resonance gate, see Fig.~\ref{fig:KAK}(d), to the transpiler. 
This ensures that at most one single-qubit pulse is needed on each qubit between each non-echoed cross-resonance $R_{ZX}$ gate.
By scaling the cross-resonance pulses we create the $R_{ZX}$ gates for arbitrary angles and therefore generalize the methods of Sec.~\ref{sec:rzx} to arbitrary gates in $SU(4)$.

We generate $R_{ZX}$-based circuits as shown in Fig.~\ref{fig:KAK}(e) for $(\alpha, \beta, \gamma)$ angles chosen at random from the Weyl chamber and measure their fidelity using process tomography with readout error mitigation.
Each $R_{ZX}$-based circuit is benchmarked against its equivalent three CNOT decomposition presented in Ref.~\cite{Vidal2004} and shown in Fig.~\ref{fig:KAK}(b).
The experiments are run on \emph{ibmq\_dublin} and \emph{ibmq\_mumbai} with 2048 shots for each circuit which we measure three times to gain statistics.
The pulse-efficient $R_{ZX}$-based decomposition of the circuits results in a significant fidelity increase for almost all angles, see Fig.~\ref{fig:SU4}.
A subset of the data is also shown in the Weyl chamber in Fig.~\ref{fig:weyl_chamber}.
The correlation between the relative error reduction and the relative schedule duration indicates that the gains in fidelity come from a better usage of the finite coherence time as the scaled cross-resonance pulses achieve the same unitary in less time.
Remarkably, these results were achieved without recalibrating any pulses.

\section{Pulse-efficient transpiler passes\label{sec:transpilation}}

The quantum circuits of an algorithm are typically expressed using generic gates such as the CNOT or controlled-phase gate and then transpiled to the hardware on which they are run~\cite{Qiskit}.
Quantum algorithms can benefit from the continuous family of gates presented in Sec.~\ref{sec:rzx} and \ref{sec:su4} if the underlying quantum circuit is either directly built from, or transpiled to, the hardware native $R_{ZX}(\theta)$ gate.
We now show how to transpile quantum circuits to a $R_{ZX}(\theta)$-based-circuit with template substitution~\cite{Iten2020}.

A template is a quantum circuit made of $|T|$ gates acting on $n_T$ qubits that compose to the identity $U_1...U_{|T|}=\mathds{1}$, see e.g. Fig.~\ref{fig:tranpiler_example}(b) and (c).
In a template substitution transpilation pass we identify a sub-set of the gates in the template $U_a...U_b$ that match those in a given quantum circuit.
Next, if a cost of the matched gates is higher than the cost of the unmatched gates in the template we replace $U_\text{match}=U_a...U_b$ with $U_\text{match}=U^\dagger_{a-1}...U^\dagger_1U_{|T|}^\dagger...U_{b+1}^\dagger$.
As cost we use a heuristic that sums the cost of each gate defined as an integer weight which is higher for two-qubit gates, details are provided in Appendix~\ref{sec:appendix_implementation}.
The complexity of the template matching algorithm on a circuit with $|C|$ gates and $n_C$ qubits is
\begin{align}
    \mathcal{O}\left(|C|^{|T|+3}|T|^{|T|+4}n_C^{n_T-1}\right),
\end{align}
i.e. exponential in the template length~\cite{Iten2020}.
We therefore create short templates where the inverse of the intended match, i.e. $U_\text{match}^\dagger$, is specified as a single gate with rules to further decompose it into $R_{ZX}$ and single-qubit gates in a subsequent transpilation pass.
In these decompositions we expose the echoed cross-resonance implementation of $R_{ZX}$ to the transpiler by writing $R_{ZX}(\theta)=X R_{ZX}(-\theta/2)XR_{ZX}(\theta/2)$.
This allows the transpiler to further simplify the single-qubit gates that would otherwise be hidden in the schedules of the two-qubit gates, as exemplified in the circuit in Fig.~\ref{fig:tranpiler_example}(e).
Finally, once the $R_{ZX}(\theta)$ gates are introduced into the quantum circuit we run a third transpilation pass to attach pulse schedules to each $R_{ZX}(\theta)$ gate built from the backend's calibrated CNOT gates following the procedure in Sec.~\ref{sec:rzx}.
The attached schedules consist of the scaled cross-resonance pulse and rotary tone without any echo.
Details on the Qiskit implementation are given in Appendix~\ref{sec:appendix_implementation}.

\section{Improving QAOA with Cartan's decomposition\label{sec:demo}}

We use the QAOA~\cite{Farhi2014, Farhi2015, Yang2017}, applied to MAXCUT, to demonstrate gains of a pulse-efficient circuit transpilation on noisy hardware.
QAOA maps a quadratic binary optimization problem with $n$ decision variables to a cost function Hamiltonian $\hat H_C=\sum_{i,j}\alpha_{i,j}Z_iZ_j$ where $\alpha_{ij}\in\mathbb{R}$ are problem dependent and $Z_i$ are Pauli $Z$ operators.
The ground state of $\hat H_C$ encodes the solution to the problem.
Next, a classical solver minimizes the energy $\braket{\psi(\boldsymbol{\beta}, \boldsymbol{\gamma})|\hat H_C |\psi(\boldsymbol{\beta}, \boldsymbol{\gamma})}$ of a trial state $\ket{\psi{(\boldsymbol{\beta}, \boldsymbol{\gamma})}}$ created by applying $p$-layers of the operator $\exp(-i\beta_k\sum_{i=0}^{j-n} X_j)\exp(-i\gamma_k\hat H_C)$ where $k=1, ..., p$ to the equal superposition of all states.

\begin{figure}[htbp!]
    \centering
    \includegraphics[width=\columnwidth, clip, trim= 4 0 4 0]{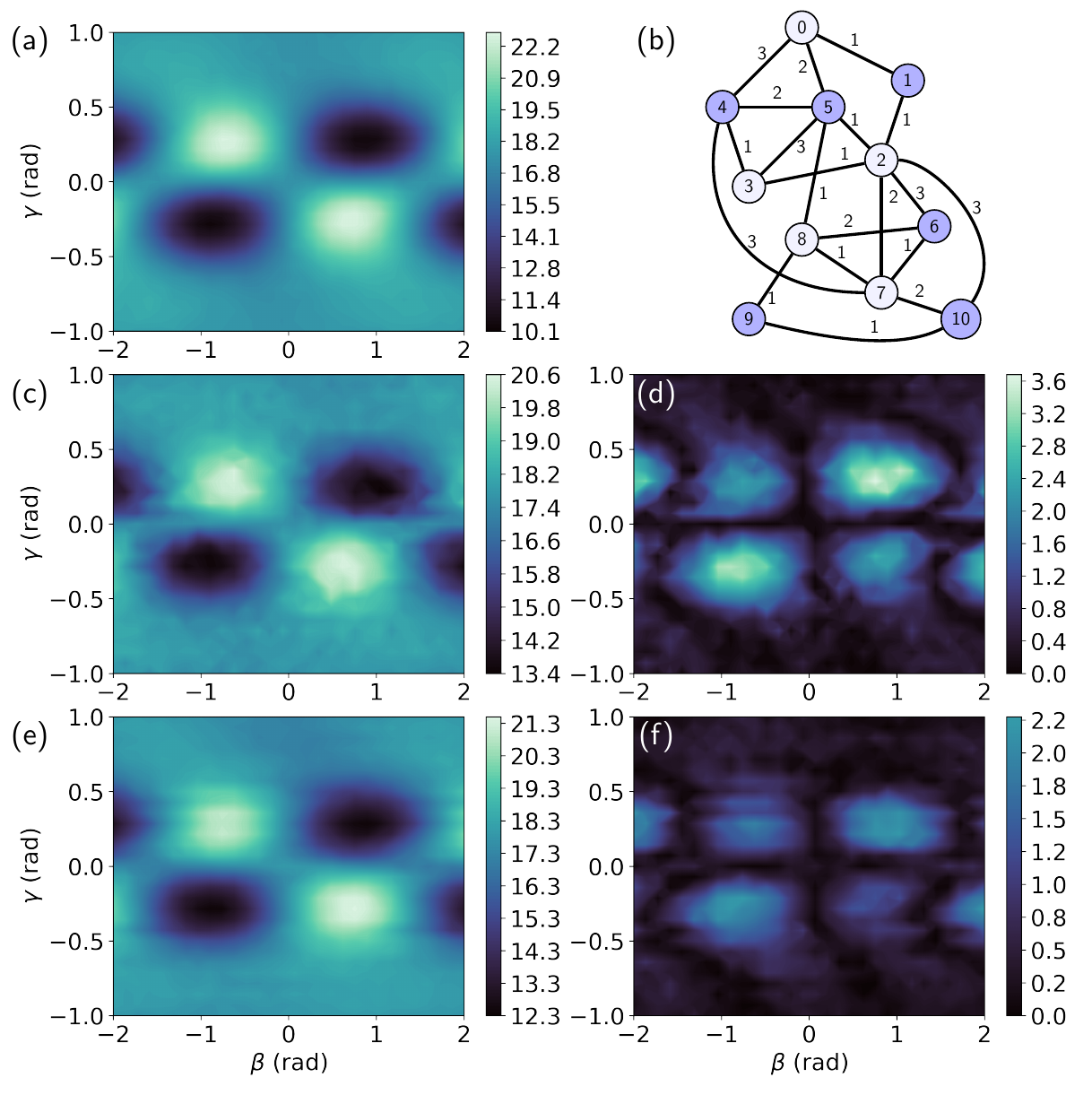}
    \caption{Depth-one QAOA energy landscape.
    (a) Noiseless simulation of the cut value, averaged over all 4096 bit-strings sampled from $\ket{\psi(\beta, \gamma)}$, obtained using the QASM simulator for the weighted graph shown in (b).
    The maximum cut, with value 28, is indicated by the color of the nodes in (b).
    Figures (c) and (e) show hardware results obtained by transpiling to CNOT gates and by using the $R_{ZX}$ pulse-efficient methodology, respectively. Figures (d) and (f) share the same color scale and show the absolute deviation from the ideal averaged cut values in figures (c) and (e), respectively.}
    \label{fig:qaoa}
\end{figure}

Implementing the operator $\exp(-i\gamma_k\hat H_C)$ requires applying the $R_{ZZ}(\theta)=\exp(-i\theta ZZ/2)$ gate on pairs of qubits.
However, to overcome the limited connectivity of superconducting qubit chips~\cite{Harrigan2021}, several $R_{ZZ}(\theta)$ gates are followed or preceded by a ${\rm SWAP}$ resulting in the unitary operator
\begin{align}
{\rm SWAP}(\theta)=
    \begin{pmatrix}
    1 & 0 & 0 & 0 \\
    0 & 0 & e^{i\theta} & 0 \\
    0 & e^{i\theta} & 0 & 0 \\
    0 & 0 & 0 & 1
    \end{pmatrix}
\end{align}
up to a global phase.
When mapped to the $KAK$ decomposition ${\rm SWAP}(\theta)$ corresponds to $\boldsymbol{k}^T=(\eta\pi/2, \eta\pi/2, \theta + \eta\pi/2))$ where $\eta=-1$ if $\theta>0$ and 1 otherwise.
This allows us to reduce the total cross-resonance duration using the methodology presented in Sec.~\ref{sec:su4}.

\begin{figure}[htbp!]
    \centering
    \includegraphics[width=\columnwidth, clip, trim= 4 0 4 0]{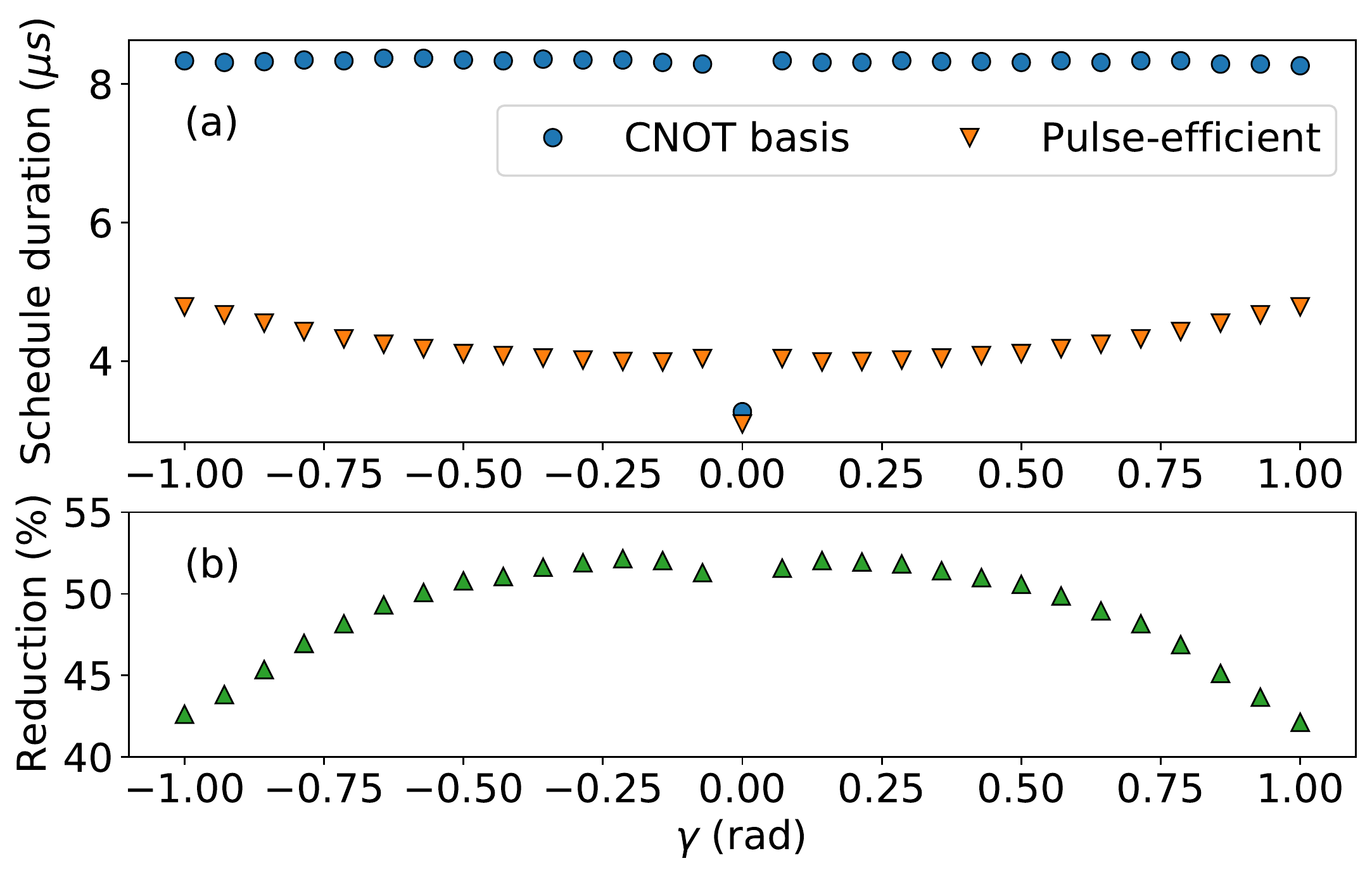}
    \caption{QAOA schedule durations. 
    (a) Duration of the scheduled quantum circuits transpiled to CNOTs with optimization level three (blue circles) and with the pulse-efficient methodology (orange triangles). 
    In both cases we removed the final measurements from the quantum circuits.
    (b) Length of the pulse efficient schedules relative to the CNOT-based schedules.}
    \label{fig:qaoa_duration}
\end{figure}

We perform a depth-one QAOA circuit for an eleven node graph, shown in Fig.~\ref{fig:qaoa}(b), built from CNOT gates.
We map the decision variables zero to ten to qubits 7, 10, 12, 15, 18, 13, 8, 11, 14, 16, 19 on \emph{ibmq\_mumbai}, respectively.
Since the graph is non-hardware-native eight $\rm SWAP$ gates are needed to implement the circuits.
In QAOA the optimal values of $(\beta, \gamma)$ are found with a classical optimizer~\cite{Bengtsson2020}.
Here, we scan $\beta$ and $\gamma$ from $\pm2~{\rm rad}$ and $\pm1~{\rm rad}$, respectively, as we submit jobs through the queue of the cloud-based IBM Quantum computers.
For each $(\beta, \gamma)$ pair we run the circuits with the noiseless QASM simulator in Qiskit, see Fig.~\ref{fig:qaoa}(a) and twice on the hardware.
The first hardware run is done using a CNOT decomposition with the Qiskit transpiler on optimization level three, see Fig.~\ref{fig:qaoa}(c) for results. 
The second run is done with the pulse-efficient circuit transpilation, see Fig.~\ref{fig:qaoa}(e) for results.
Here, we first perform the template substitution with the $R_{ZZ}(\theta)$ and ${\rm SWAP}(\theta)$ templates, shown in Fig.~\ref{fig:tranpiler_example}(b), (c) and Appendix~\ref{sec:appendix_implementation} for further details.
A second transpilation pass then exposes the $R_{ZX}(\theta)$ gates to which we attach pulse schedules in a third transpilation pass following Sections \ref{sec:rzx} -- \ref{sec:transpilation}.
In each case we measure 4096 shots. 
The pulse-efficient circuits produce less noisy average cut values, compare Fig.~\ref{fig:qaoa}(c) with (e), and have a lower absolute deviation from the noiseless simulation than the circuits transpiled to CNOT gates, compare Fig.~\ref{fig:qaoa}(d) with (f).
The maximum error in the cut value averaged over the sampled bit-strings is reduced by 38\% from 3.65 to 2.26.
We attribute the increased quality of the results to the decrease in total cross-resonance time and the fact that the pulse-efficient transpilation keeps the number of single-qubit pulses to a minimum.
In total, we observe a reduction in total schedule duration ranging from 42\% to 52\% depending on $\gamma$ when using the pulse efficient transpilation methodology, see Fig.~\ref{fig:qaoa_duration}.
Since the schedule duration of  $R_{ZZ}(\gamma\alpha_{i,j})$ and ${\rm SWAP}(\gamma\alpha_{i,j})$ decreases and increases as $\gamma$ decreases, respectively, we observe a non-monotonous reduction in the schedule duration of the QAOA circuit as a function of $\gamma$.

\section{Discussion and Conclusion}

The results in Sec.~\ref{sec:rzx} and \ref{sec:su4} showed that by scaling cross-resonance gates we can automatically create a continuous family of gates which implements $SU(4)$.
These scaled gates typically have shorter pulse schedules and higher fidelities than the digital CNOT implementation.
This fidelity is limited by coherence, imperfections in the initial calibration, and non-linear effects.
Crucially, the resulting gate-tailored pulse schedules do not require additional calibration and can therefore be automatically generated by the transpiler.
Transpilation passes, as discussed in Sec.~\ref{sec:transpilation}, can be leveraged to identify and attach the scaled pulse schedules to the gates in a quantum circuit.
Furthermore, exposing the echo in the cross-resonance gate to the transpiler allows further simplifications of the single-qubit gates.
We used this pulse-efficient transpilation methodology to reduce errors in an eleven-qubit depth-one QAOA.

Scaled gates are particularly appealing for Trotter based applications, as shown in Ref.~\cite{Stenger2020}, and could therefore benefit quantum simulations~\cite{Tornow2020}.
Future work may also include scaling direct cross-resonance gates~\cite{Jurcevic2021} and benchmarking their impact on Quantum Volume~\cite{Cross2019}.
Methods to interpolate pulse parameters based on a set of reference $R_{ZX}(\theta)$ gates, calibrated at a few reference angles $\theta$, might also improve the gate fidelity and help deal with non-linearities between the rotation angle $\theta$ and pulse parameters.
For variational algorithms, such as the variational quantum eigensolver, the scaled $SU(4)$ gates may allow for better results due to the shorted schedules while still being robust to some unitary errors such as angle errors~\cite{Colless2018, Egger2019}.

We believe that the methods presented in our work will help users of noisy quantum hardware to reap the benefits of pulse-level control without having to know its intricacies. 
This can improve the quality of a broad class of quantum applications running on noisy quantum hardware.

\section{Acknowledgments}

The authors acknowledge use of the IBM Quantum devices for this work.
The authors also thank L. Capelluto, N. Kanazawa, N. Bronn, T. Itoko and E. Pritchett for insightful discussions and S. Woerner for a careful read of the manuscript.

\bibliography{references}

\newpage

\appendix

\section{Qiskit implementation\label{sec:appendix_implementation}}

Quantum circuits often have repeating sub-circuits with different parameters.
For instance, QAOA circuits include many $R_{ZZ}(\gamma\alpha_{ij})$ and ${\rm SWAP}(\gamma\alpha_{ij})$ gates where $\gamma$ is one of the  variational parameters and the $\{\alpha_{ij}\}$ depend on the problem instance.
We therefore need parametric templates when running the template substitution algorithm.

\begin{figure}[htbp!]
    \centering
    \includegraphics[width=\columnwidth, clip, trim=0 0 10 0]{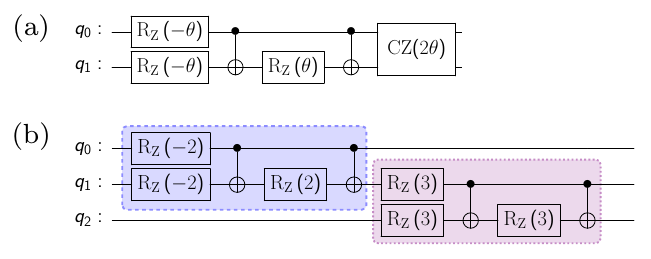}
    \caption{(a) Parametric template of a controlled-$Z$ gate. (b) Circuit on which the template matching is run. The dashed blue and dotted purple boxes indicate potential matches based on circuit instruction names and qubits.}
    \label{fig:appendix_template}
\end{figure}

We extended the Qiskit implementation of Ref.~\cite{Iten2020} to parametric templates.
To avoid a symbolic description of the unitary matrix of each gate we first match gates by qubits and name.
This is however not sufficient to create a valid match since, for example, the parametric template in Fig.~\ref{fig:appendix_template}(a) produces two tentative matches on the circuit in Fig.~\ref{fig:appendix_template}(b).
We therefore form a system of equations based on the tentative match.
If this system of equations accepts a solution the match is valid.
For example, the tentative match in Fig.~\ref{fig:appendix_template}(b), indicated by the dashed blue box, results in the system of equations
\begin{align}
\begin{cases}
    -\theta=-2\\
    -\theta=-2\\
    \phantom{-}\theta=2
\end{cases}
\end{align}
which accepts the solution $\theta=2$ and is therefore valid.
However, the second tentative match, highlighted by the dotted purple box, results in the system of equations
\begin{align}
\begin{cases}
    -\theta=3\\
    -\theta=3\\
    \phantom{-}\theta=3
\end{cases}
\end{align}
which has no solution and is therefore not valid.

\begin{figure}
    \centering
    \includegraphics[width=0.95\columnwidth, clip, trim=133 455 240 122]{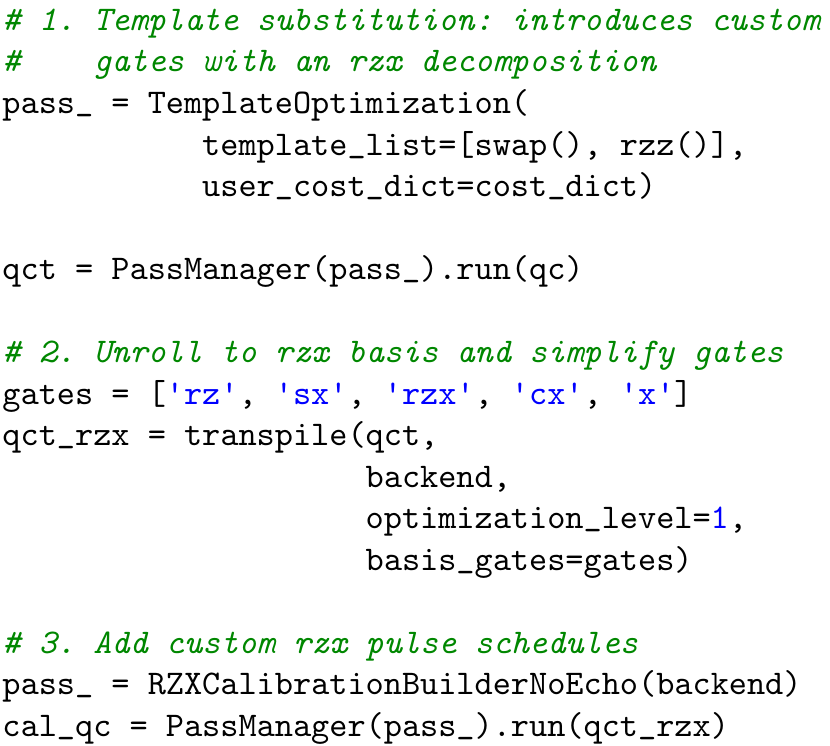}
    \caption{Example of a circuit transpilation that achieves a pulse-efficient circuit transpilation.}
    \label{fig:code_example}
\end{figure}

\begin{figure}[htbp!]
    \centering
    \includegraphics[width=\columnwidth, clip, trim= 0 5 40 20]{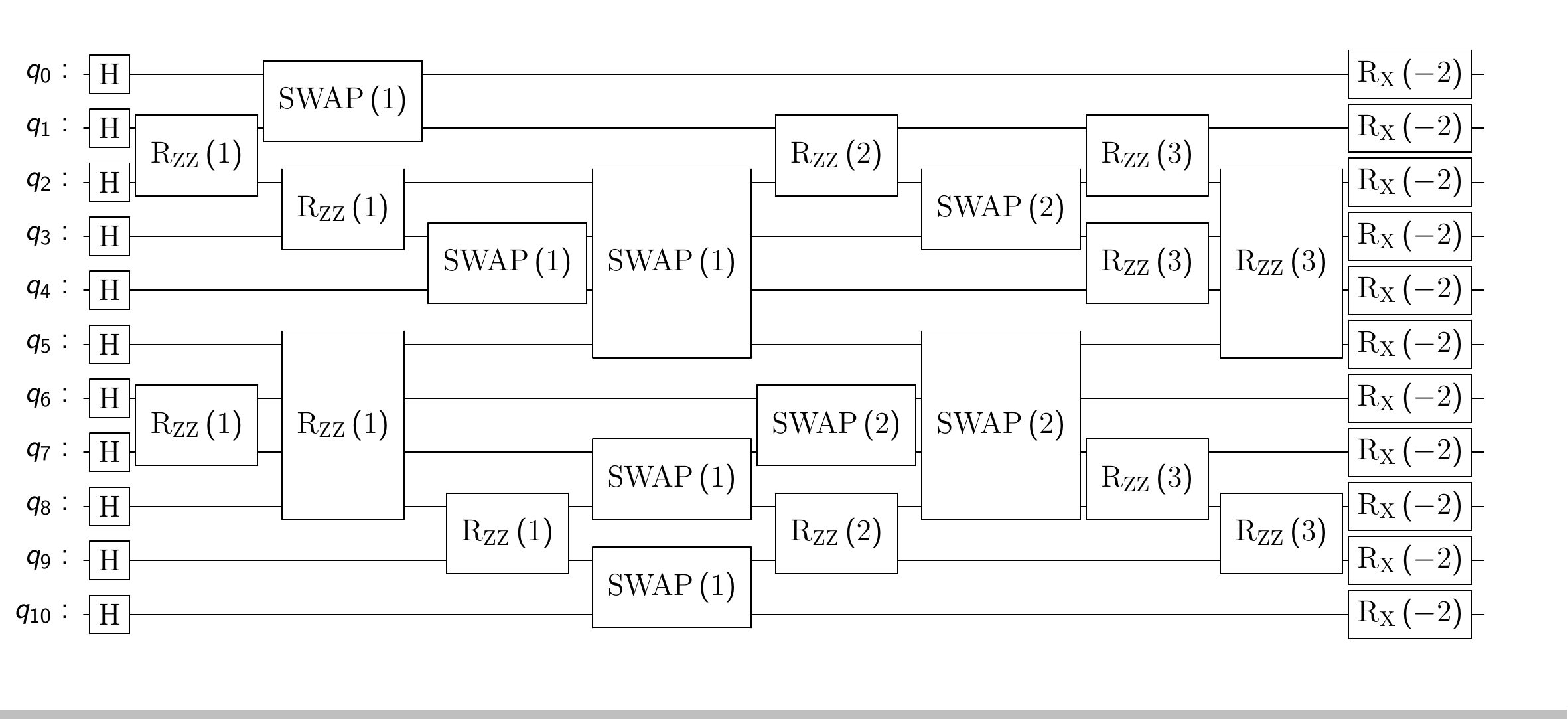}
    \caption{11 qubit QAOA circuit with $\gamma=1$ and $\beta=-2$ after the template substitution. The final measurement instructions have been omitted.}
    \label{fig:qaoa_circ_1}
\end{figure}

We achieve a pulse-efficient circuit transpilation with Qiskit by using three transpilation steps shown in Fig.~\ref{fig:code_example}.
First, the \texttt{TemplateOptimization} transpilation pass is applied with the \texttt{SWAP} and \texttt{rzz} templates as shown in Fig.~\ref{fig:tranpiler_example}(b) and (c) of the main text.
The next step, a standard transpiler pass with a low optimization level, i.e. one, exposes the \texttt{rzx} definition of the gates in the matched templates.
Finally, the \texttt{RZXCalibrationBuilderNoEcho} class scales the pulses of the cross-resonance gates and attaches them to the $R_{ZX}(\theta)$ gates in the circuit.
Figure~\ref{fig:qaoa_circ_1} exemplifies the result of the first transpilation pass applied to the QAOA circuit in Sec.~\ref{sec:demo}.

The template optimization pass requires a cost dictionary to determine if it is favourable to replace the matched gates $U_\text{match}=U_a...U_b$ from a template with the Hermitian conjugate of the remaining part of the template $U^\dagger_{a-1}...U^\dagger_1U_{|T|}^\dagger...U_{b+1}^\dagger$.
The cost dictionary has gates as keys and their cost as value.
The cost of $U_a...U_b$ is the sum of the costs of each individual gate $U_a$ to $U_b$.
We used the cost dictionary
\texttt{\{'sx': 1, 'x': 1, 'rz': 0, 'cx': 2, 'rzz': 0, 'swap': 6, 'phase\_swap': 0\}}
which assigns a zero cost to the \texttt{rzz} and \texttt{phase\_swap} gates which correspond to the pulse-efficient implementation of $R_{ZZ}(\theta)$ and ${\rm SWAP}(\theta)$.
Single-qubit gates have unit cost except for \texttt{rz} which is implemented with virtual $Z$-rotations. 
The CNOT gate, i.e. \texttt{cx}, and the standard $\rm SWAP$ gate, i.e. \texttt{swap} have costs two and six, respectively.
This cost dictionary ensures that the template substitution will include $R_{ZZ}(\theta)$ and ${\rm SWAP}(\theta)$ in the case of a match.
Future work could improve this heuristic cost dictionary either by using the fidelity of the gates (if this metric is available) or the duration of the underlying pulse schedules as cost.

\section{Properties of the Quantum devices and additional data\label{sec:appendix_additional_data}}

\begin{figure*}[htbp!]
    \centering
    \includegraphics[width=2\columnwidth]{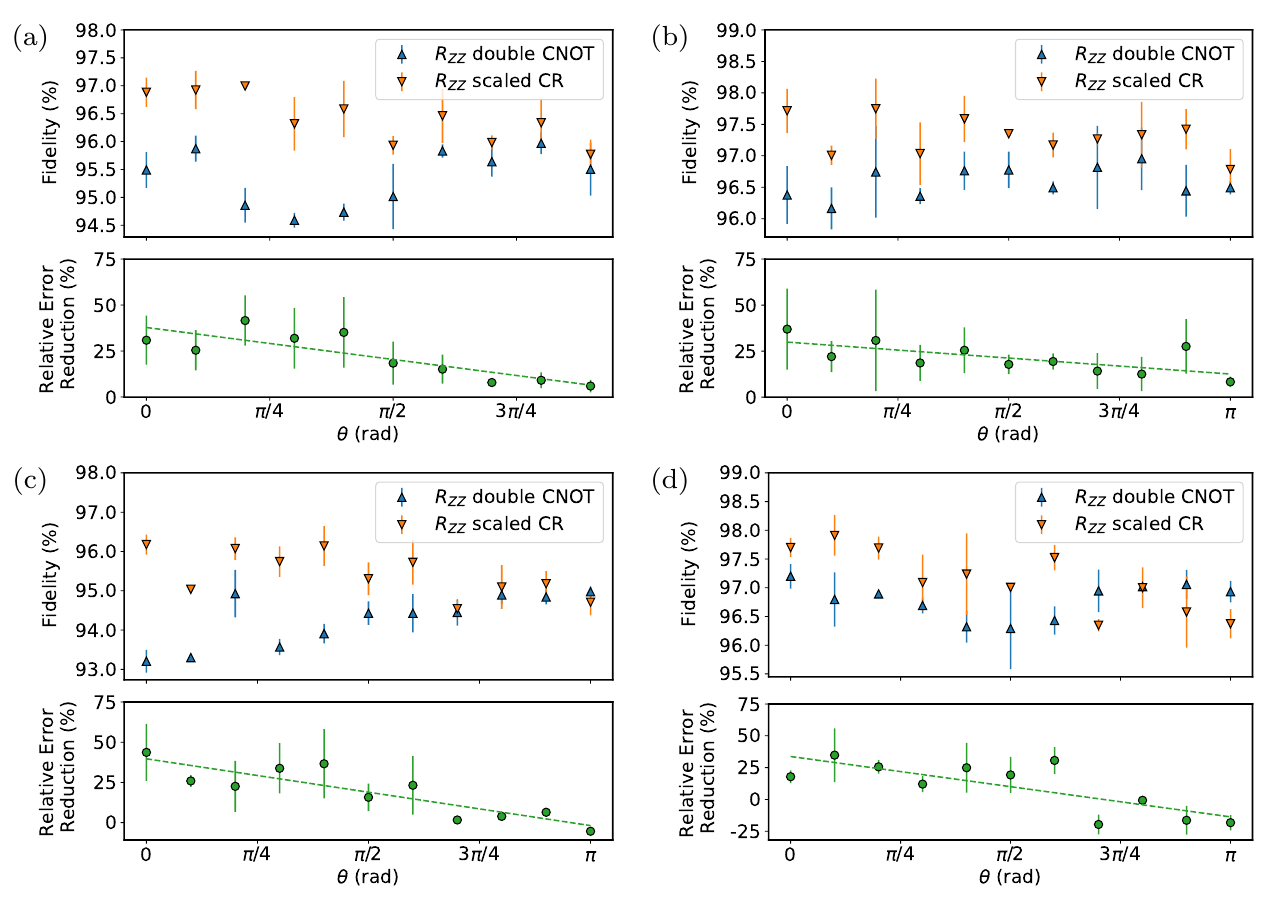}
    \caption{Gate fidelity measured with quantum process tomography and relative error reduction for the $ZZ$-gate as a function of $\theta$ on \emph{ibmq\_mumbai} (qubits 16 and 19), \emph{ibmq\_montreal} (qubits 14 and 16), \emph{ibmq\_paris} (qubits 1 and 2) and \emph{ibmq\_guadalupe} (qubits 7 and 10). (a-d, top) Process fidelities for the $R_{ZZ}$ double CNOT (blue up-triangles) and scaled CR (orange down-triangles) circuit implementation. (a-d, bottom) Relative error reduction calculated from the fidelities.\label{fig:fidelities_zz_appendix}}
\end{figure*}

Since the qubit coherence times as well as the $\rm CNOT$ gate duration and error mainly limit the fidelity of the scaled cross-resonance gates we list their values for the qubits and devices we experimented with in Tab.~\ref{tab:devices}.
To illustrate that scaling imperfect cross-resonance gates improves the gate fidelity we measured the process fidelity on several IBM Quantum devices and qubit pairs.
In almost all measurements the scaled gates have a higher fidelity than the double $\rm CNOT$ benchmark and the relative error reduction increases as the schedule duration decreases, see Fig.~\ref{fig:fidelities_zz_appendix}.

\begin{table}[htbp!]
    \centering
    \caption{Summary of the properties of the $\rm CNOT$ gates and coherence times for the qubits used to benchmark the performance of the scaled cross-resonance gates.}
    \label{tab:devices}
    \begin{tabular}{l r r r r} \hline\hline
        & \multicolumn{2}{c}{CNOT} & & \\
        Device & error & duration &
        $\quad T_1$-times & $\quad T_2$-times \\
        & (\%) & $(\mathrm{ns})$ & $(\mu\mathrm{s})$ & $(\mu\mathrm{s})$ \\ \hline\hline
        \emph{ibmq\_mumbai} \\
        q1, q2 & 1.27 & 739 & $102,\,157$ & $34,\,228$ \\
        q16, q19 & 0.84 & 754 & $84,\,141$ & $105,\,132$ \\ \hline
        \emph{ibmq\_paris} \\
        q1, q2 & 1.70 & 597 & $66,\,92$ & $82,\,128$ \\
        q13, q14 & 1.28 & 434 & $100,\,23$ & $27,\,33$ \\
        q18, q15 & 5.36 & 448 & $86,\,74$ & $103,\,50$ \\
        q18, q17 & 1.76 & 725 & $41,\,71$ & $94,\,157$ \\ \hline
        \emph{ibmq\_dublin} \\
        q1, q2 & 0.76 & 540 & $110,\,103$ & $174,\,89$ \\
        q3, q2 & 0.83 & 370 & $78,\,96$ & $100,\,83$ \\ \hline
        \emph{ibmq\_montreal} \\
        q14, q16 & 0.88 & 356 & $97,\,87$ & $97,\,52$ \\ \hline
        \emph{ibmq\_guadalupe} \\
        q7, q10 & $0.61$ & $299$ & $99,\,68$ & $153,\,90$ \\ \hline\hline
    \end{tabular}
\end{table}

\begin{figure*}[htbp!]
    \centering
     \includegraphics[width=2\columnwidth]{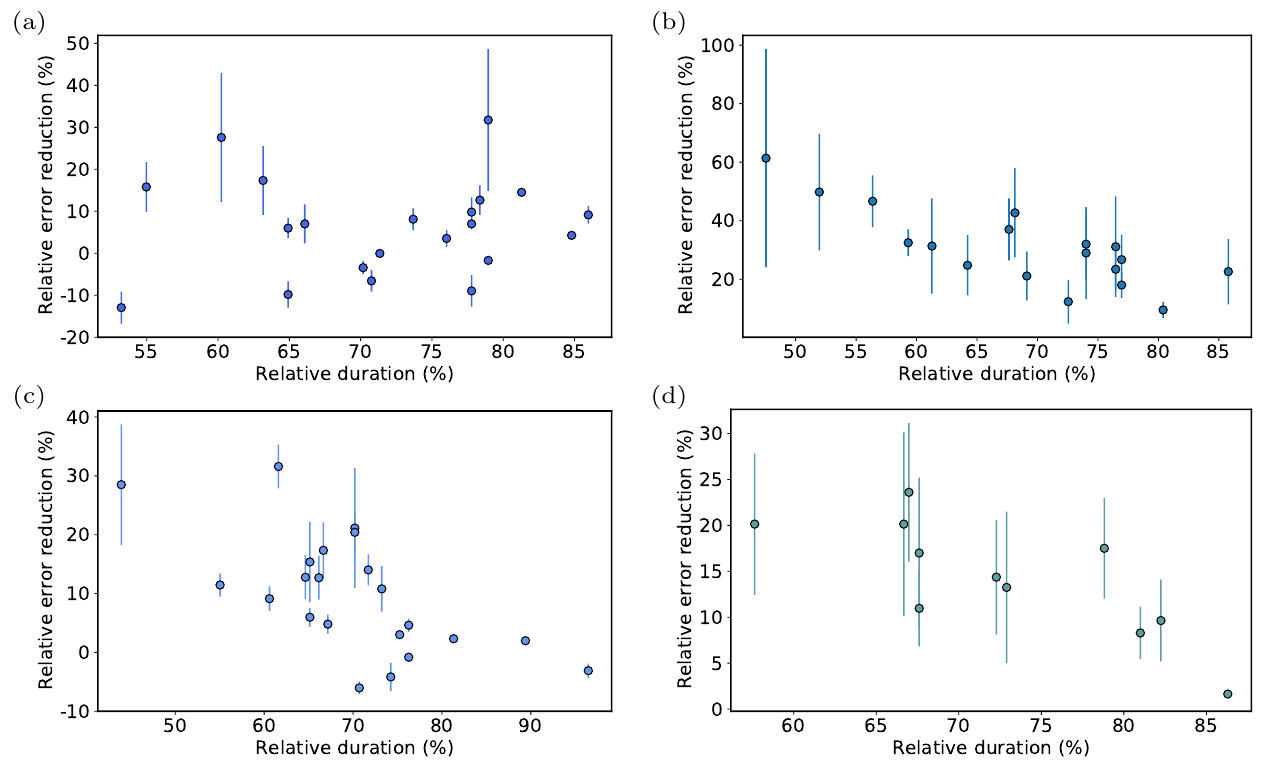}
    \caption{Quantum Process Tomography results for random angles in the Weyl chamber on (a) \emph{ibmq\_dublin} (qubits 3 and 2), (b)  \emph{ibmq\_paris} (qubits 18 and 15), (c) \emph{ibmq\_paris} (qubits 13 and 14) and (d) \emph{ibmq\_paris} (qubits 18 and 17).}
    \label{fig:SU4_appendix}
\end{figure*}

Fig.~\ref{fig:SU4_appendix} shows additional quantum process tomography results for  Cartan-decomposed circuits chosen at random in the Weyl chamber.
The experiments were performed on \emph{ibmq\_dublin}, see Fig.~\ref{fig:SU4_appendix}(a), and \emph{ibmq\_paris} using different qubit pairs, see Fig.~\ref{fig:SU4_appendix}(b) -- (d).
For almost all angles the relative error reduction is positive which demonstrates the advantage of a hardware-native, scaled cross-resonance gate based circuit implementation.

\section{Theoretical coherence limit\label{sec:appendix_fidelity_limit}}

In Sections~\ref{sec:rzx} we compared the relative decrease in gate error to the coherence limit on the average gate error $\mathcal{E}$.
This limit is implemented in Qiskit Ignis for two qubits $a$ and $b$ as $\mathcal{E}=\frac{3}{4}\left(1-u_1-u_2\right)$ where
\begin{align}
    u_1 =& \frac{1}{15}\left(e^{-t/T_{1, a}}+e^{-t/T_{1,b}}+e^{-t/T_{1,a}-t/T_{1,b}}\right), \\
    u_2 =& \frac{2}{15}\left(e^{-t/T_{2,b}}+e^{-t/T_{2,b}-t/T_{1,a}}+e^{-t/T_{2,a}}\right. \\ \notag
    &+\left.e^{-t/T_{2,a}-t/T_{1,b}}+2e^{-t/T_{2,a}-t/T_{2,b}}\right).
\end{align}
The derivation of this limit is discussed in more detail in Appendix G of Ref.~\cite{Sundaresan2020}.
Here, $t$ is the gate duration while $T_{1, a}$ and $T_{2, a}$ are the $T_1$ and $T_2$ times for qubit $a$.
Since the process fidelity and the average gate fidelity are linearly related~\cite{Horodecki1999, Magesan2011} we compare the relative error reduction in the measured process fidelity with the theoretical relative error reduction in $\mathcal{E}$.


\newpage

\end{document}